\documentclass [12pt]{article}%
\usepackage{amsmath}
\usepackage{amssymb}
\usepackage{epsfig}
\usepackage{latexsym}
\usepackage{amsfonts}
\usepackage{graphicx}%
\usepackage{varioref}
\usepackage{ifthen}
\newcommand{\mpi}{\mu_\pi} \newcommand{\mpio}{m_{0 \pi}}
\newcommand{\mrho}{M_\rho} \newcommand{\mrhoo}{m_{0 \rho}}
\newcommand{\rpp}{{\rho\pi\pi}}
\newcommand{\mn}{{\mu\nu}}
\newcommand{\xint}{{\int_0^1 d{x_1} \int_0^{1-x_1}d{x_2}}}
\newcommand{\jpi}{{\phi^{*} \overleftrightarrow{\partial_\mu} \phi}}
\newcounter{bracketlevel}
\newcommand{\bracket}[2][big]
{{
\addtocounter{bracketlevel}{1}
\ifthenelse{\equal{#1}{big}}{
\ifthenelse{\value{bracketlevel}=1}{{\left( #2 \right)}}{
 \ifthenelse{\value{bracketlevel}=2}{{\left[ #2 \right]}}{
  \ifthenelse{\value{bracketlevel}=3}{{\left\{ #2 \right\}}}{
  \left( #2 \right)
  }}}
}{
\ifthenelse{\value{bracketlevel}=1}{{( #2 )}}{
 \ifthenelse{\value{bracketlevel}=2}{{[ #2 ]}}{
  \ifthenelse{\value{bracketlevel}=3}{{\{ #2 \}}}{
    ( #2 )
  }}}
}
\addtocounter{bracketlevel}{-1}
}}

\begin{document}
\parindent 0mm \setlength{\parskip}{\baselineskip}
\thispagestyle{empty}
\pagenumbering{arabic}
\setcounter{page}{0} \mbox{ }
\rightline{UCT-TP-267/07}\newline\rightline{(revised) August 2007}
\newline%
\begin{center}
{\large \textbf{Pion form factor in the Kroll-Lee-Zumino model}}
{\LARGE \footnote{{\LARGE {\footnotesize Supported in part by FONDECYT 1051067 and 7050125, and Centro de Estudios Subatomicos (Chile),
and NRF (South Africa).}}}}
\end{center}
\vspace{.1cm}
\begin{center}
\textbf{Cesareo A. Dominguez}$^{(a)}$, \textbf{Juan I. Jottar}$^{(b),(c)}$, 
\textbf{Marcelo Loewe}$^{(b)}$, \\
\textbf{Bernard Willers}$^{(a)}$ 
\end{center}
\begin{center}
$^{(a)}$Centre for Theoretical Physics and Astrophysics,University of
Cape Town, Rondebosch 7700, South Africa

$^{(b)}$Facultad de F\'{i}sica, Pontificia Universidad Cat\'{o}lica de Chile, Casilla 306, Santiago 22, Chile

$^{(c)}$ Department of Physics, University of Illinois, Urbana-Champaign, \\
IL\, 61801 3080, USA\end{center}

\vspace{0.3cm}
\begin{center}
\textbf{Abstract}
\end{center}
The renormalizable Abelian quantum field theory model of Kroll, Lee, and Zumino is used to compute the one-loop vertex corrections to the tree-level, Vector Meson Dominance (VMD) pion form factor. These corrections, together with the known one-loop vacuum polarization contribution, lead to  a substantial improvement over VMD. The resulting pion form factor in the space-like region is in excellent agreement with data in the whole range of accessible momentum transfers. The time-like form factor, known to reproduce the Gounaris-Sakurai formula at and near the rho-meson peak, is unaffected by the vertex correction at order $\cal{O}$$(g_\rpp^2)$.
\noindent 


\newpage
\bigskip
\noindent
The renormalizable Abelian quantum field theory of charged pions, and  massive neutral vector mesons, proposed long ago by Kroll, Lee, and Zumino (KLZ) \cite{KLZ}, provides a rigorous theoretical justification for the Vector Meson Dominance (VMD) ansatz \cite{VMD}. The fact that in this model the  neutral vector mesons are coupled only to conserved currents ensures  renormalizability \cite{KLZ},\cite{Hees}. A very interesting phenomenological application of this model was made some time ago by Gale and  Kapusta  \cite{GK} who computed the rho-meson self energy to one-loop order. When this result is used in the VMD expression for the pion form factor, there follows the well known Gounaris-Sakurai formula \cite{GS}-\cite{tau} in the time-like region at and near the rho-meson pole. We find this quite intriguing. That an empirical fit formula such as this should follow from  the KLZ Lagrangian may be hinting at additional unexpected properties of this model. In this note we explore this possibility by computing the vertex diagram, i.e. the one loop correction to the strong coupling constant in the framework of the KLZ model. This correction is of the same order in the coupling as the one loop vacuum polarization. After regularization and renormalization, and in conjunction with the VMD expression for the pion form factor, this vertex correction, together with the vacuum polarization contribution, leads to an excellent agreement between theory and experimental data in the space-like region. The parameter free result (masses and couplings are known from experiment)  constitutes a substantial improvement over naive (tree-level) VMD. In fact, the resulting chi-squared per degree of freedom is close to unity, while the one from tree-level VMD is about five times bigger. 
Predictions in the time-like region are shown to be unaffected by the vertex correction. In fact, the combination of vacuum polarization and vertex corrections in this region turns out to be of higher order in the coupling. Clearly, 
since the KLZ model involves a strong coupling, the perturbative expansion could be questioned, and the next-to-leading (one-loop) contributions need not be smaller than the leading term.
However, this is not the case with the KLZ model. In fact, the  relatively small $\rho\pi\pi$ coupling ($g_{\rho\pi\pi} \simeq 5$) is accompanied by the large loop suppression factor $1/(4 \pi)^2$, so that the one-loop contributions remain reasonable corrections to the leading order tree-level term. At higher orders, we expect higher powers of this  suppression factor from loop integrations. However, a detailed next-to-next-to leading order calculation is beyond the scope of this work.\\

We begin by introducing the KLZ Lagrangian,
\begin{equation}
\mathcal{L}_{KLZ} = \partial_\mu \phi \; \partial^\mu \phi^* -  m_\pi^2 \;\phi \;\phi^* - \tfrac{1}{4}\; \rho_\mn \;\rho^\mn + \tfrac{1}{2}\; m_\rho^2\; \rho_\mu \;\rho^\mu 
+g_{\rho\pi\pi} \rho_\mu J^\mu_\pi\;,
\end{equation}
where $\rho_\mu$ is a vector field describing the $\rho^0$ meson ($\partial_\mu \rho^\mu = 0$), $\phi$ is a complex pseudo-scalar field describing the $\pi^\pm$ mesons, $\rho_\mn$ is the usual field strength tensor, and $J^\mu_\pi$ is the $\pi^\pm$ current, i.e.

\begin{equation}
\rho_\mn  = \partial_\mu \rho_\nu - \partial_\nu \rho_\mu \; ,
\end{equation}

\begin{equation}
J^\mu_\pi  = i \jpi \; .
\end{equation}

Omitted from Eq.(1)  is an additional term of higher order in the coupling, of the form $g_\rpp^2 \;\rho_\mu\; \rho^\mu \;\phi \;\phi^*$, which is not relevant to the present work.
\begin{figure}[ht]
\begin{center}
\includegraphics[width=60mm]{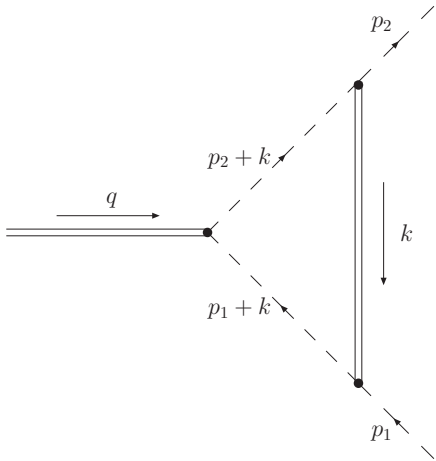}
\caption{Vertex function kinematics.}
\end{center}
\end{figure}

In Fig. 1 we define the vertex function kinematics. Using the Feynman propagator for the $\rho$-meson \cite{Hees},\cite{Quigg} and in $d$-dimensions, the unrenormalized vertex is given by
\begin{multline}
\widetilde{\Gamma}^{(1)\mu}_\rpp ({p_1}, {p_2}, q^2) = g_\rpp^3 \bracket{\mu^3}^{2 - \frac{d}{2}} \\ 
\times \int \frac{d^dk}{\bracket{2 \pi}^d}
\frac{({p_1} + {p_2} + 2k)^\mu \, (2 {p_1} + k) \cdot (2 {p_2} + k)} 
{[({p_1} + k)^2 - m_\pi^2 + i \varepsilon] [({p_2} + k)^2 - m_\pi^2 + i \varepsilon] (k^2 - m_\rho^2 + i \varepsilon)} \;.
\label{vertex0}
\end{multline}

Using the Feynman identity for the propagators, Eq.(\ref{vertex0}) can be rewritten as

\begin{eqnarray}
\widetilde{\Gamma}^{(1)\mu}_\rpp ({p_1}, {p_2}, q^2) &=& 2  g_\rpp^3 \left( \mu^3 \right)^{2 - \frac{d}{2}} \xint  \int \frac{d^dl}{\bracket{2 \pi}^d} \frac{(2l-2s+p_1+p_2)^\mu}{(l^2 - \Delta(q^2) + i \varepsilon)^3} \nonumber \\ [.3cm]
& \times & \Big[ l^2 + (2 p_1 - s) \cdot (2 p_2 - s) + 2 l \cdot (p_1 + p_2 - s) \Big] \;, \label{vertex1}
\end{eqnarray}

where the new momentum $l$ is defined as
\begin{equation}
l = k + \bracket{{x_1}{p_1} + {x_1}{p_2}} \;,
\label{def:l}
\end{equation}
the parameter $s$ is
\begin{equation}
s = \bracket{{x_1}{p_1} + {x_1}{p_2}} \;,
\label{def:s}
\end{equation}

and the real variable $\Delta(q^2)$ is given by
\begin{equation}
\Delta(q^2) = m_\pi^2 \bracket{x_1 + x_2}^2 + m_\rho^2 \bracket{1 - x_1 - x_2} - {x_1}{x_2}q^2 \;.
\label{def:delta}
\end{equation}

Introducing the integrals
\begin{equation}
I_s = \int \frac{d^dl}{(2 \pi)^{d}} \frac{l^{2s}}{(l^2 - \Delta(q^2) + i \varepsilon)^3} \; ,
\label{def:Is}
\end{equation}
and the functions
\begin{equation}
f_1 (x_1, x_2)  =  \Big[ m_\pi^2 (x_1 + x_2 - 2 )^2 - q^2(x_1x_2 - x_1 - x_2 + 2) \Big] I_0 + \Big[ 1 + \tfrac{4}{d} \Big] I_1\\ \;,
\end{equation}
and
\begin{equation}
f_2 (x_1, x_2)  =  \Big[ m_\pi^2 (x_1 + x_2 - 2 )^2 - q^2(x_1x_2 - x_1 - x_2 + 2) \Big] I_0 + \Big[ 1 + \tfrac{2}{d} \Big] I_1\\ \;,
\end{equation}

the vertex function becomes
\begin{eqnarray}
\widetilde{\Gamma}^{(1)\mu}_\rpp ({p_1}, {p_2}, q^2) & =& 2 \, g_\rpp^3 \left( \mu^3 \right)^{2 - \frac{d}{2}}(p_1 + p_2)^\mu \xint \left[  f_1(x_1, x_2) \right. \nonumber \\ [.3cm]
&-& \left . 2 x_1 f_2 (x_1, x_2) \right] 
 = {\Gamma}^{(0)\mu}_\rpp (p_1, p_2) \;G(q^2) \;,
\label{def:Gamma1}
\end{eqnarray}

where
\begin{equation}
\Gamma^{(0)\mu}_\rpp (p_1, p_2) = i g_\rpp \mu^{(2 - \frac{d}{2})}(p_1+p_2)^\mu \;, 
\end{equation}

is the \emph{tree level} vertex in $d$ dimensions, and 

\begin{eqnarray}
G(q^2) & \equiv &  g_\rpp^2 \left( \mu^2 \right)^{(2 - \frac{d}{2})} \frac{2}{i} \xint [f_1(x_1, x_2) - 2 x_1 f_2(x_1, x_2)] \nonumber \\
[.3cm]
&= & g_\rpp^2 \left( \mu^2 \right)^{(2 - \frac{d}{2})} \frac{2}{i} \xint \left\{ \left[ (1- 2x_1) + \frac{4}{d} (1 - x_1) \right] I_1 \right. \nonumber \\ [.3cm]
&+& \left.  (1 - 2x_1) \phantom{\frac{1}{1}}\Big[ m_\pi^2 (x_1 + x_2 - 2 )^2 - q^2(x_1x_2 - x_1 - x_2 + 2) \Big] I_0 \right\}.
\label{def:G1}
\end{eqnarray} 
 
Evaluating the integrals $I_0$ and $I_1$ in dimensional regularization leads to

\begin{eqnarray}
G(q^2) &=&  -2 \;\frac{g_\rpp^2}{(4 \pi)^2} \left( \mu^2 \right)^{(2 - \frac{d}{2})} \xint \left\{ ( 2 - 3 x_1 ) \left[ \frac{2}{\varepsilon} - \ln \left( \frac{\Delta(q^2)}{\mu^2} \right) \right. \right.\nonumber \\
[.3cm]
 &-& \left. \left. \frac{1}{2} - \gamma + \ln \left( 4 \pi \right) \right] 
  + \frac{(1 - 2x_1)}{2 \, \Delta} \left[ m_\pi^2 (x_1 + x_2 - 2 )^2 \right. \right. \nonumber \\
[.3cm]
  &-& \left. \left. q^2(x_1x_2 - x_1 - x_2 + 2) \right] + \phantom{\frac{1}{1}} \mathcal{O}(\varepsilon)  \right\} \;.
\end{eqnarray}

Separating the terms involving divergences and constants from the rest of the expression this equation can be rewritten as

\begin{equation}
G(q^2) = \widetilde{G}(q^2) + A \left[ \frac{2}{\varepsilon} -\frac{1}{2} - \gamma + \ln (4 \pi) \right] + \mathcal{O}(\varepsilon) \;,
\label{G:A}
\end{equation}
where $\widetilde{G}(q^2)$ is the $\frac{1}{\varepsilon}$ divergence free function of $q^2$, i.e.

\begin{multline}
\widetilde{G}(q^2) =  - 2 \frac{g^2_\rpp}{(4 \pi)^2} \xint \left\{(2 - 3x_1) \ln \left( \frac{\Delta(q^2)}{\mu^2} \right) \right.\\
\left. + \left( \frac{1 - 2 x_1}{2 \Delta(q^2)} \right) \Big{[} m_\pi^2 (x_1 + x_2 - 2)^2 - q^2(x_1 x_2 - x_1 - x_2 + 2) \Big{]} \right\} \;.
\label{Gsquig}
\end{multline}

The factor $A$ in Eq.(16) is an integral over $x_1$ and $x_2$, but does not depend on $q^2$. As a result, it is a constant that will be cancelled during renormalization and there is no need to calculate it explicitly. It is easy to show that this vertex function  develops an imaginary part above the two-pion threshold ($q^2 \geq 4 \, m_\pi^2$).\\ 

 The renormalization programme is quite standard \cite{Hees}-\cite{GK}, as sketeched in the following. First, the KLZ Lagrangian Eq.(1) is understood as involving bare quantities (pion and rho-meson fields, masses and coupling) denoted with a subscript $0$. Next, a rescaling  is performed

\begin{equation}
\begin{split}
\phi_0 & = Z_\phi^{\frac{1}{2}} \phi \\
\rho^0_\mu & = Z_\rho^{\frac{1}{2}} \rho_\mu \;,
\end{split}
\end{equation}

where $Z_\phi$ and $Z_\rho$ are the renormalization constants associated with each field. Inserting these into the Lagrangian yields

\begin{eqnarray}
\mathcal{L}_0 &=& Z_\phi \partial_\mu \phi \partial^\mu \phi^* - Z_\phi \mpio^2 \phi^* \phi - Z_\rho \tfrac{1}{4} \rho_\mn \rho^\mn + Z_\rho \tfrac{1}{2} \mrhoo^2 \rho_\mu \rho^\mu \nonumber \\ 
&+& i Z_\phi Z_\rho^\frac{1}{2} g_{0\rpp} \rho^\mu \phi^* \overleftrightarrow{\partial_\mu} \phi \;.
\end{eqnarray}

Now define

\begin{equation}
\begin{aligned}
\delta Z_\phi & = Z_\phi - 1 \qquad & \qquad  \delta Z_\rho & = Z_\rho - 1 \\
\delta \mpi^2 & = \mpio^2 Z_\phi - \mpi^2 & \delta \mrho^2 & = \mrhoo^2 Z_\rho - \mrho^2 \\
g_\rpp Z_g & = g_{0\rpp} Z_\phi Z_\rho^\frac{1}{2} & \delta Z_g & = Z_g - 1 \;,\\
\end{aligned}
\label{def:renormZ}
\end{equation}

where $\mpi$, $\mrho$, and $g_\rpp$  are the physically measured mass of $\pi^\pm$, mass of the $\rho^0$, and the $\rho\pi\pi$ coupling, respectively. 
The full Lagrangian now separates into terms involving only physically measurable quantities and a set of counter terms,viz.

\begin{equation}
\mathcal{L}_0  = \mathcal{L} + \Delta \mathcal{L} \;, \end{equation}

\begin{equation}
\mathcal{L} = \partial_\mu \phi \partial^\mu \phi^* - \mpi^2 \phi^* \phi - \tfrac{1}{4} \rho_\mn \rho^\mn + \tfrac{1}{2} \mrho^2 \rho_\mu \rho^\mu + i g_{\rpp} \rho^\mu \phi^* \overleftrightarrow{\partial_\mu} \phi 
\end{equation}

\begin{eqnarray}
\Delta \mathcal{L} &=& \delta Z_\phi \partial_\mu \phi \partial^\mu \phi^* - \delta \mpi^2 \phi^* \phi - \tfrac{1}{4} \delta Z_\rho  \rho_\mn \rho^\mn + \tfrac{1}{2} \delta \mrho^2 \rho_\mu \rho^\mu \nonumber \\[.3cm]
&+& i \delta Z_g g_{\rpp} \rho^\mu \phi^* \overleftrightarrow{\partial_\mu} \phi \;.
\end{eqnarray}

The five $\delta$ coefficients in the counter terms require the definition of five renormalization conditions. There are four homogeneous conditions which define the position of the pion and rho-meson propagator poles, and their unit residues, which determine $\delta \mpi^2$, $\delta \mrho^2$, $\delta Z_\phi$ and $\delta Z_\rho$. These have no practical bearing on the vertex function, hence we concentrate on the renormalization condition for the latter. 
For reasons to become clear below, wee choose for the vertex function the renormalization point $q^2=0$ and the condition

\begin{equation}
\Gamma^{(1)\mu}_\rpp (p_1,p_2,q^2 = 0)  = \Gamma^{(0)}_\rpp(p_1,p_2) \;.
\end{equation}

The bare vertex function Eq.(12) is now replaced by the renormalized one according to

\begin{equation}
\begin{split}
\Gamma^{(1)\mu}_\rpp (p_1, p_2, q^2) & = \Gamma^{(0)\mu}_\rpp (p_1, p_2) + \widetilde{\Gamma}^{(1)\mu}_\rpp (p_1, p_2) \\[.3cm]
& = \Gamma^{(0)\mu}_\rpp (p_1, p_2) \left[ 1 + G(q^2) + \delta Z_g \right]\;,
\end{split}
\end{equation}

which, using Eq.(16) it becomes

\begin{equation}
\Gamma^{(1)\mu}_\rpp (p_1, p_2, q^2) = \Gamma^{(0)\mu}_\rpp (p_1, p_2) \left\{ 1 + \widetilde{G}(q^2) + A \left[ \frac{2}{\varepsilon} - \frac{1}{2} - \gamma + \ln (4 \pi) \right] +  \delta Z_g \right\}.
\end{equation}

The renormalization condition Eq.(24) implies

\begin{equation}
\delta Z_g = - \widetilde{G}(0) -   A \left[ \frac{2}{\varepsilon} - \frac{1}{2} - \gamma + \ln (4 \pi) \right] \;,
\end{equation}

yielding the renormalised vertex function

\begin{equation}
\begin{split}
\Gamma^{(1)\mu}_\rpp (p_1, p_2, q^2) & = \Gamma^{(0)\mu}_\rpp (p_1, p_2) \left[ 1 + \widetilde{G}(q^2) - \widetilde{G}(0) \right] \\[.3cm]
& = i (p_1 + p_2)^\mu g_\rpp \left[ 1 + \widetilde{G}(q^2) - \widetilde{G}(0) \right] \;.
\label{renorm:vertex:0}
\end{split}
\end{equation}

We have chosen to  renormalize  the vertex at the off-shell point $q^2 = 0$, where $G(q^2)$ is purely real,
to make use of the known normalization of the pion form factor, $F_\pi(0) = 1$. This allows us to obtain the renormalized vertex function involving no additional constants, i.e.

\begin{eqnarray}
G(q^2) - G(0) &=& - 2\; \frac{g_\rpp^2}{(4 \pi)^2} \xint \left\{(2 - 3x_1)  \ln \left( \frac{\Delta(q^2)}{\Delta(0)} \right) \right. \nonumber \\
[.3cm]
&+&  \left. \frac{(1 - 2 x_1)}{2}  \left[ \mpi^2 (x_1 + x_2 - 2)^2 \left(\frac{1}{\Delta(q^2)} - \frac{1}{\Delta(0)}\right)
 \right. \right. \nonumber  \\ [.3cm]
&-& \left. \left. \frac{q^2}{\Delta(q^2)} (x_1 x_2 - x_1 - x_2 +2) \right] \right\}\;,
\end{eqnarray}

and where $g_\rpp \equiv g_\rpp(q^2=0)$. The pion form factor in VMD at tree level is given by the well known expression

\begin{equation}
F_\pi(q^2)|_{\mbox{VMD}} = \frac{g_\rpp}{f_\rho}\; \frac{M_\rho^2}{M_\rho^2 - q^2} \;.
\end{equation}
 
The  pion form factor including the one-loop vertex correction
at order $\cal{O}$$(g_\rpp^2)$ can then be written as

\begin{equation}
F_\pi(q^2)|_{\mbox{vertex}} = \frac{g_\rpp}{f_\rho}\; \frac{M_\rho^2}{M_\rho^2 - q^2} \;[1+ G(q^2) - G(0)] ,
\end{equation}

where $f_\rho = 4.97 \pm 0.07$ \cite{PDG}, and from universality and $F_\pi(0) = 1$ it follows that $g_\rpp(0) = f_\rho$. Hence, the one-loop vertex correction generates an additional momentum dependence in the form factor; this turns out to be a smooth monotonically decreasing function of $q^2$. Numerically, it is a reasonable correction to the tree-level result due to the relatively mild coupling, and to the strong suppression factor $1/(4 \pi)^2$ from the loop integration.

\begin{figure}[ht]
\begin{center}
\includegraphics[width=\columnwidth]{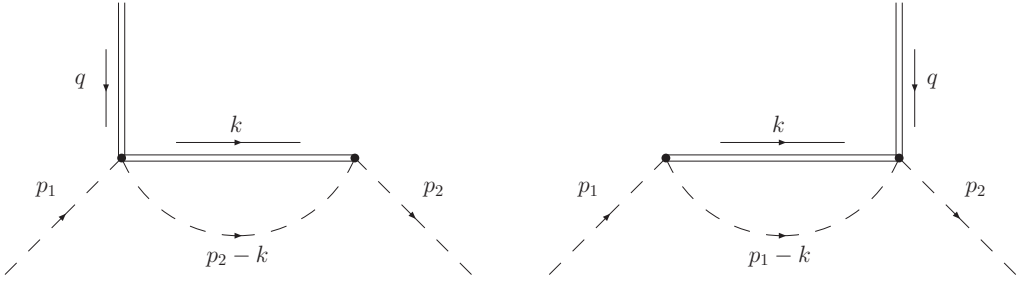}
\caption{Seagull, $q^2$-independent corrections to tree level at order $\cal{O}$$(g_\rpp^2)$.}
\end{center}
\end{figure}

In addition to the vertex correction there are two seagull-type corrections to  tree-level at the same order $\cal{O}$$(g_\rpp^2)$, as illustrated in Fig. 2. It is easy to show, though, that after regularization and renormalization these diagrams do not contribute to the form factor. In fact, as they are $q^2$-independent, they cancel after subtraction at $q^2=0$.  Nevertheless, these diagrams do contribute to the renormalization constants (of the masses and fields but not the coupling), and are essential to ensure gauge invariance of the vertex correction.
To complete the calculation of the pion form factor at order $\cal{O}$$(g_\rpp^2)$ one has to include the vacuum polarization contributions from the diagrams shown in Fig. 3. The calculation of these diagrams is standard in scalar electrodynamics with a massive photon  \cite{Quigg} and it has been discussed in \cite{GK}, the result being

\begin{eqnarray}
\Pi(q^2)|_{\mbox{vac}} &=& \frac{1}{3}\; \frac{g_\rpp^2}{(4 \pi)^2}
\; \;q^2 \; \;\Big{(}1 - 4\; \frac{\mu_\pi^2}{q^2}\Big{)}^{3/2} \; \left[
\ln \Bigg| \frac{\sqrt{(1 - 4 \;\mu_\pi^2/q^2)} + 1}{\sqrt{(1 - 4 \;\mu_\pi^2/q^2)} - 1} \Bigg| \right. \nonumber \\[.3cm]
&-& \left. i \; \pi\; \theta(q^2 - 4 \mu_\pi^2) \phantom{\frac{1}{1}} \right] + A \; q^2 + B \; ,
\end{eqnarray}

where the constants $A$ and $B$ are
\begin{equation}
A = - \frac{1}{3} \; \frac{g_\rpp^2}{(4 \pi)^2}\; \Bigg{[} 8\; \frac{\mu_\pi^2}{M_\rho^2} + \Big{(}1 -  4 \; \frac{\mu_\pi^2}{M_\rho^2}\Big{)}^{3/2} \; \ln \Bigg| \frac{\sqrt{(1 - 4 \;\mu_\pi^2/M_\rho^2)} + 1}{\sqrt{(1 - 4\; \mu_\pi^2/M_\rho^2)} - 1} \Bigg| \Bigg{]} \;,
\end{equation}

\begin{equation}
B = \Pi(0)|_{\mbox{vac}} = \frac{8}{3} \; \frac{g_\rpp^2}{(4 \pi)^2}\; \mu_\pi^2 \;.
\end{equation}

The tadpole contribution in Fig.3, proportional to $g_{\mu \nu}$, cancels an identical piece from the first diagram, rendering the result purely transverse. Adding the vacuum polarization  to the vertex contribution gives the complete correction to the VMD pion form factor at order $\cal{O}$$(g_\rpp^2)$

\begin{equation}
F_\pi(q^2) = \frac{M_\rho^2 + \Pi(0)|_{\mbox{vac}}}{M_\rho^2 - q^2 + \Pi(q^2)|_{\mbox{vac}}} + \frac{M_\rho^2}{M_\rho^2 - q^2} \Big[ G(q^2) - G(0)\Big] \;,
\end{equation} 

where $\Pi(q^2)|_{\mbox{vac}}$ is given in Eq.(32), and $G(q^2)$ is obtained after performing a  numerical integration in Eq.(29).
\begin{figure}[t]
\begin{center}
\includegraphics[width=\columnwidth]{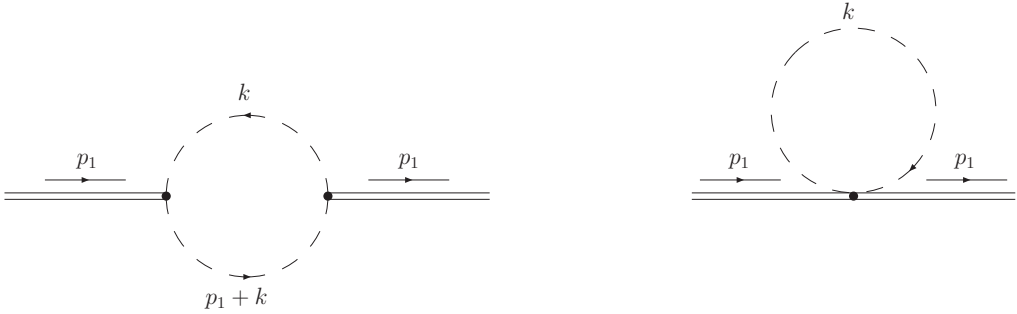}
\caption{Vacuum polarization contributions at order $\cal{O}$$(g_\rpp^2)$}
\end{center}
\end{figure}
\begin{figure}[ht]
\begin{center}
\includegraphics[width=\columnwidth]{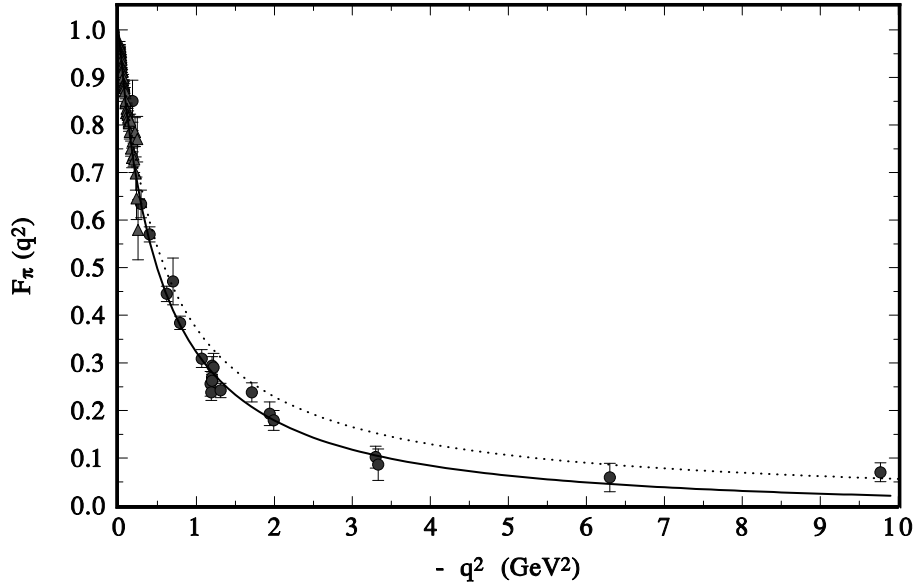}
\caption{Pion form factor data together with the KLZ prediction, Eq.(35) (solid line), and the tree-level  VMD result (dotted line).}
\end{center}
\end{figure}
This result contains no free parameters, as the masses and the coupling are known from experiment.
Notice that the vacuum polarization correction is not included in the second term above, as it would make this term of order   $\cal{O}$$(g^4)$. Hence, the vertex correction does not affect the form factor in the time-like region, where it becomes the Gounaris-Sakurai formula near the rho-meson peak. In fact, from the definition of the hadronic width \cite{PIL}: $\Gamma_\rho = - (1/M_\rho)\;\,Im \; \Pi(M_\rho^2)  $, where $\Gamma_\rho \equiv \Gamma_\rho(M_\rho^2)$, and from Eq.(32) there follows

\begin{equation}
\Gamma_\rho = \frac{g_{\rho\pi\pi}^2}{48 \pi}\; \frac{1}{M_\rho^2} \; (M_\rho^2 - 4 \, \mu_\pi^2)^{\frac{3}{2}} \;,
\end{equation}

which is the standard kinematical relation between width and coupling of a vector and two pseudoscalar particles \cite{PIL}. Notice that this results follows automatically in the KLZ model, i.e. it has not been imposed as a constraint. Near the rho-meson peak, where $\Pi(s)$ is largely purely imaginary,  the s-dependent width which follows from Eqs. (32) and (36) is

\begin{equation}
\Gamma_\rho(s)|_{KLZ} = \frac{M_\rho\, \Gamma_\rho}{\sqrt{s}} \Big[\frac{s - 4\, \mu_\pi^2}{M_\rho^2 - 4\, \mu_\pi^2}\Big]^{\frac{3}{2}} \;,
\end{equation}

which is precisely the momentum dependent Gounaris-Sakurai width \cite{PIL}.
This is known to provide an excellent fit to the data in this region \cite{tau}. \\

\begin{figure}[h]
\begin{center}
\includegraphics[width=\columnwidth]{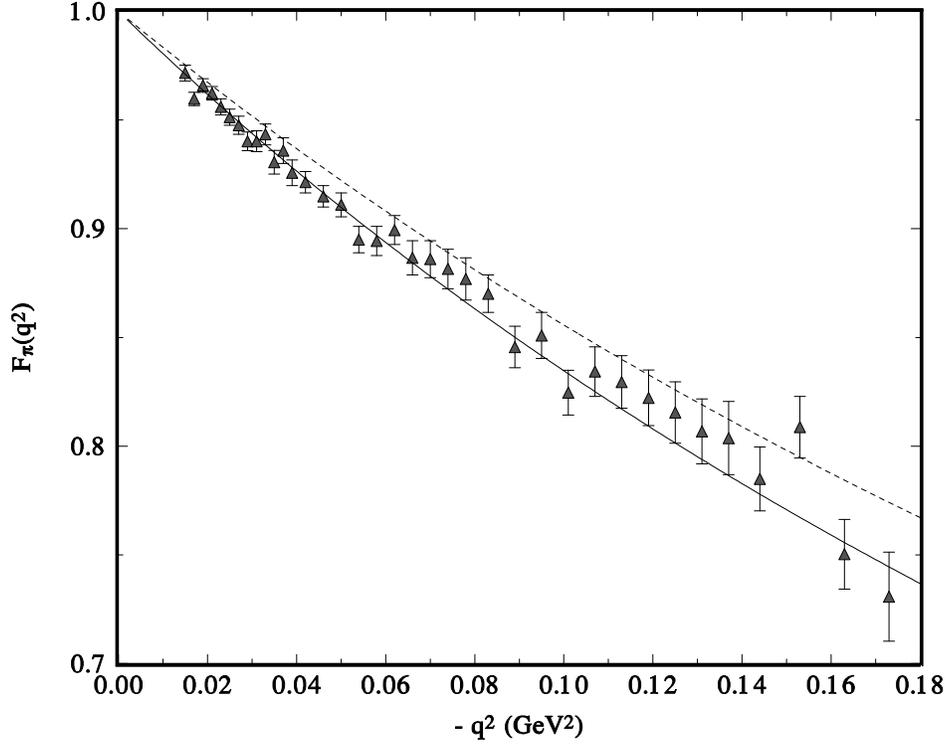}
\caption{Pion form factor data at low momenta, together with the KLZ prediction, Eq.(22) (solid line), and the tree-level  VMD result (dotted line).}
\end{center}
\end{figure}

Turning to the space-like region, the form factor Eq. (35) is plotted in Fig.4 (solid line) together with the experimental data \cite{data} and the reference prediction from tree-level VMD (dotted curve). The latter provides a poor fit to the data as evidenced from the resulting chi-square per degrees of freedom $\chi_F^2 = 5.0$, while Eq.(35) gives the optimal value $\chi_F^2 = 1.1$. In addition, the mean-square radius of the pion obtained from Eq.(35) is $<r^2_\pi> = 0.40 \;\mbox{fm}^2$, to be compared with a similar result from
tree-level VMD  $<r^2_\pi> = 6/M_\rho^2 = 0.39 \;\mbox{fm}^2$, and the experimental value
  $<r^2_\pi> = 0.439 \; \pm \;0.008\; \mbox{fm}^2$. 
For reasons of scale, it is difficult to appreciate visually the agreement of the KLZ form factor with the data at small momenta. This is the region where the errors are smaller, hence the region that counts the most towards achieving a low chi-squared. In Fig. 5 we show the data in this region together with the KLZ form factor (solid curve) and tree-level VMD (dotted line). This kind of excellent agreement between theory and experiment is comparable to that obtained from Dual Large $N_c$ QCD ($QCD_{\infty}$) \cite{DNQCD} which gives $\chi_F^2 = 1.2$. $QCD_{\infty}$ is a Dual Resonance Model (Veneziano) realization of QCD in the limit of an infinite number of colours. In this limit QCD is solvable and the hadronic spectrum consists of an infinite number of zero-width resonances \cite{NC}. The masses and couplings of these states remain unspecified, though, so that one needs a model to fix them. Dual- $QCD_{\infty}$, after unitarization in the time-like region, bears some resemblance to KLZ in the sense of generating a correction to naive VMD, in this case single rho-dominance. The infinite set of vector meson radial excitations in Dual- $QCD_{\infty}$ correspond to the loop corrections in KLZ.
But then, contrary to KLZ, $QCD_{\infty}$ involves one free parameter in the space-like region. Unitarization of  the $QCD_{\infty}$ pion form factor in the time-like region gives a reasonable result at and around the rho-meson peak. However, the  KLZ form factor stands aside as it reproduces the Gounaris-Sakurai formula in this region.\\
  
In summary, the KLZ one-loop level contributions to the pion form factor turn out to be reasonable corrections to the leading order result. This is in spite of KLZ being a strong interaction theory. This is due to the relatively mild coupling ($g_\rpp \simeq 5$), together with a large loop suppression factor ($(1/4 \pi)^2)$, as seen from Eqs. (15), (17) and (29). Increasing powers of this suppression factor are expected at higher orders in perturbation theory. An explicit two-loop calculation, though, is beyond the scope of the present work. The parameter-free prediction for the pion form factor leads to excellent agreement with data for both space-like and time-like momenta. In view of its renormalizability, plus the successful predictions for the pion form factor, we wish to argue the case for the KLZ model to be considered as a viable tool to analyze $\pi \pi$ dynamics \cite{pipi}. One should keep in mind, though, that a good part of that dynamics (involving charged rho-mesons) would remain excluded if one were to insist on renormalizability.

{\bf Acknowledgments}\\
The authors  wish to thank Marco Aurelio Diaz, Gary Tupper, Nello Paver, and Karl Schilcher for valuable discussions.\\


\begin{thebibliography}{99}
\bibitem{KLZ} N.M. Kroll, T.D. Lee, B. Zumino, Phys. Rev. 175 (1967) 1376; J.H. Lowenstein, B. Schroer,Phys. Rev. D 6 (1972) 1553.
\bibitem{VMD} J.J. Sakurai, Ann. Phys. (N.Y.) 11 (1960) 1; {\it ibid.} Currents and Mesons, University of Chicago Press (1969).
\bibitem{Hees} H. van Hees, hep-th/0305076 (unpublished); H. Ruegg, M. Ruiz-Altaba, Int. J. Mod. Phys. A 19 (2004) 3265.
\bibitem{GK} C. Gale, J. Kapusta, Nucl. Phys. B 357 (1991) 65.
\bibitem{GS} G. Gounaris, J.J. Sakurai, Phys. Rev. Lett. 21 (1968) 244; see also M. Gourdin, Phys. Rep. 11 C (1974) 29.
\bibitem{tau} M. Davier, A. H\"{o}cker, Z. Zhang, Rev. Mod. Phys. 78 (2006) 1043.
\bibitem{Quigg} C. Quigg, Gauge Theories of Strong, Weak, and Electromagnetic Interactions, Benjamin (1983).
\bibitem{PDG} Particle Data Group,  J. Phys. G 33 (2006) 1. 
\bibitem{PIL} H.M. Pilkuhn, {\it Relativistic Particle Physics}, Springer Verlag (1979).
\bibitem{data} C.J. Bebek {\it et al.}, Phys. Rev. D 17 (1978) 1693; S.R. Amendolia {\it et al.}, Nucl. Phys. B 277 (1986) 168;, J. Volmer {\it et al.}, Phys. Rev. Lett. 86 (2001) 1713.
\bibitem{DNQCD} C.A. Dominguez, Phys. Lett.B 512 (2001) 331.
\bibitem{NC} G. 't Hooft, Nucl. Phys. B 72 (1974) 461;
 E. Witten, Nucl. Phys. B 79 (1979) 57.
\bibitem{pipi} Low energy $\pi-\pi$ scattering in KLZ beyond leading order is currently being analyzed in: C.A. Dominguez, M. Loewe, K. Schilcher (in preparation).
\end{thebibliography}
\end{document}